\documentclass[doublecol]{epl2} 

\title{Orbital Compass Model as an Itinerant Electron System }
\shorttitle{
Orbital Compass Model as an Itinerant Electron System 
} 

\author{J. Nasu and S. Ishihara}
\shortauthor{J. Nasu and S. Ishihara}

\institute{                    
   Department of Physics, Tohoku University - Sendai 980-8578, Japan
}
\pacs{75.25.Dk}{Orbital, charge, and other orders, including coupling of these orders}

\abstract{
Two-dimensional orbital compass model is studied as an interacting itinerant electron model. A Hubbard-type tight-binding model, from which the orbital compass model is derived in the strong coupling limit, is identified. This model is analyzed by the random-phase approximation (RPA) and the self-consistent RPA methods from the weak coupling. Anisotropy for the orbital fluctuation in the momentum space is qualitatively changed by the on-site Coulomb interaction. This result is explained by the fact that the dominant fluctuation is changed from the intra-band nesting to the inter-band one by increasing the interaction. 
}

\usepackage{amsfonts}
\usepackage{amsmath}	
\usepackage{bm}	

\newcommand{\means}[1]{\langle#1\rangle}

\usepackage{mathrsfs} 

\def\PRB{Phys.\ Rev.\ B}
\def\PRL{Phys.\ Rev.\ Lett.}

\def\JMMM{J.~Mag.~Mag.~Mat.}

\def\EPL{Europhys.\ Lett.}

\begin{document}

\maketitle

\section{Introduction}
Orbital degree of freedom is one of the central and not yet solved issues in correlated electron systems\cite{Tokura,book}. This degree of freedom is recognized as a key to elucidate several exotic phenomena in solid state physics such as colossal magneto-resistance effect\cite{Tokura99}, iron-based high-$T_c$ superconductivity\cite{Nakayama}, and so on. Essence of the orbital is a directional nature; electron motions, and electron-electron interactions depend on directions in a crystal lattice~\cite{Khomskii}. This characteristic brings about several non-trivial results such as a macroscopic number of degeneracy in the orbital configurations, the dimensional reduction in the effective interaction and others~\cite{Feiner,Nussinov,Ishihara97}. 

One of the well studied orbital models is the Kugel-Khomskii model\cite{Kugel}, which describes the interaction between the localized ions with spin and orbital degrees of freedom in the nearest neighbor (NN) sites. The interactions on each NN bond are represented by products of the Heisenberg-type spin part and the orbital part, which is described by the pseudo-spin (PS) operators, ${\bm T}_i$.  In the latter, the orbital interactions explicitly depend on directions of the bonds in a crystal lattice. This model is obtained from the Hubbard-type itinerant electron model with orbital degenracy in the strong coupling limit. It is recognized that the magnetic and elastic properties in several transition-metal compounds, e.g KCuF$_3$ and LaMnO$_3$, are well reproduced by this model Hamiltonian. 

Another well known and simple orbital model is the orbital compass model, where the orbital degree of freedom is only taken into account, instead of the spin-orbital entanglement in the Kugel-Khomskii model. In this Hamiltonian, the component of the PS operator concerned in an interaction explicitly depends on the direction of a bond, e.g., only the $x$ component $T_i^x$ for the interaction along the $x$ direction. In particular, the compass model in the two-dimensional square lattice has been studied so far intensively and extensively from the view points of the orbital order in a Mott insulator\cite{Jackeli} as well as qubits in the quantum computer\cite{Doucot}, and the topological order~\cite{Nussinov08}. Some new concepts, such as the directional order~\cite{Mishra}, a non-trivial dilution effect~\cite{Tanaka}, and the generalized Elitzur's theorem~ \cite{Batista}, have been proposed through the theoretical examinations. 

In this Letter, we study the two-dimensional orbital compass model from the view point of the interacting itinerant electron system. We identify the Hubbard-type model from which the orbital compass model is derived in the strong coupling limit. The model Hamiltonian is analyzed by the random-phase approximation (RPA) method and the self-consistent (SC) RPA method. It is found that anisotropy in the orbital fluctuation strongly depends on the on-site Coulomb interaction. The results are interpreted by the intra-band and inter-band nestings. Relations between the present results and those in the original compass model are discussed.

\section{Model}
We start from the orbital compass model on a square lattice in the $x$-$z$ plane defined by
\begin{align} 
 H_{\rm Compass}=J\sum_{<ij>_x}T_i^x T_j^x +J\sum_{<ij>_z}T_i^z T_j^z,\label{eq:1}
\end{align}
where $J$ is the exchange constant and $<ij>_l$ represents the NN sites along the $l(=x, z)$ direction. The  PS  operator  $\bm{T}_i$  with  a  magnitude  of  1/2  describes a doubly-degenerate orbital degree of freedom. This is given by 
$
 \bm{T}_i=\frac{1}{2}\sum_{\gamma\gamma'}c_{i\gamma}^\dagger
 {\bm \sigma}_{\gamma \gamma'}c_{i\gamma'},
$
 where $c_{i\gamma}$ is  the annihilation operator for a spin-less fermion with orbital $\gamma(=a, b)$ at site $i$, and ${\bm \sigma}$ are the Pauli matrices.

Here we introduce the Hubbard-type spin-less fermion model from which the orbital compass model in eq.~(\ref{eq:1}) is derived by the perturbational procedure in the strong coupling limit. This is given by 
\begin{align}
H=H_t+H_U,  \label{eq:2}
\end{align}
with
\begin{align}
H_t= 
-\sum_{<ij>_l \gamma \gamma'}
\left (c_{i\gamma}^\dagger t_l^{\gamma
\gamma'}c_{j\gamma'}+{\rm H.c.} \right )
-\mu \sum_{i \gamma} c_{i\gamma}^\dagger c_{i\gamma}, \label{eq:2a}
\end{align} 
and 
\begin{align}
H_U= U\sum_i n_{ia}n_{ib},\label{eq:2b}
\end{align} 
where $n_{i\gamma}( \equiv  c_{i\gamma}^\dagger  c_{i\gamma}$) is the number operator. In the first term in eq.~(\ref{eq:2}), $t_l^{\gamma \gamma'}$ is the transfer integral between the $\gamma$ and $\gamma'$ orbitals along the direction $l$, and $\mu$ is the chemical potential. In the second term, $U$ is the on-site Coulomb interaction between the spin-less fermions with different orbitals. The matrix elements of the transfer integral are determined to reproduce the compass model in eq.~(\ref{eq:1}) by the second-order perturbation with respect to the transfer integral. First we consider the exchange interaction along the $z$ direction, i.e. $T_i^z T_j^z$. This Ising-type interaction requires a condition that the transfer integral along this direction is finite only for one of the two orbitals, i.e. 
\begin{align}
t_z=\frac{1}{2}t(1+\sigma^z) , 
\label{eq:tz}
\end{align}
where $t$ is a constant.  In the same way, the transfer integral along the $x$ direction is given as 
\begin{align}
t_x=\frac{1}{2}t(1+\sigma^x) .  
\label{eq:tx}
\end{align}
We consider the so-called half-filled case where a number of fermions is equal to that of the lattice sites. 

The transfer integrals introduced above give the following energy bands in the momentum space, 
\begin{align}
 H_t=\sum_{\bm{k}\gamma\gamma'}c_{\bm{k}\gamma}^\dagger  (\varepsilon_{\bm{k}}^{\gamma\gamma'}
-\mu)c_{\bm{k}\gamma'} , 
\end{align}
with 
\begin{align}
 \varepsilon_{\bm{k}}=-t(1+\sigma^x)\cos k_x-t(1+\sigma^z)\cos k_z . 
 \label{eq:bareband}
\end{align}
We take a lattice constant as a unit of length. It is worth to note that these band-dispersions are generalized as 
\begin{align}
\varepsilon_{\bm{k}}(\theta)=-t\left [ 1+\sigma(\theta) \right ]\cos k_x-t(1+\sigma^z)\cos k_z , 
\label{sigmaeta}
\end{align}
where $\sigma(\theta)=\sigma^z \cos \theta + \sigma^x \sin \theta$ with a real number $\theta$. It is obvious that the bands with $\theta=\pi/2$ are reduced to those in the compass model. The parameter values $\theta=2\pi/3$ and $\pi$ correspond to the models where the doubly degenerate $(3z^2-r^2/x^2-y^2)$ orbitals and the $(yz/xy)$ ones are introduced at each site in a square lattice, respectively~\cite{Cincio,Daghofer}. By diagonalizing the matrix, we obtain the two bands 
\begin{align}
 H_t=\sum_{\eta=(\pm) \bm{k}}
 (E_{\bm{k}}^\eta-\mu)
 d_{\bm{k}\eta}^\dagger d_{\bm{k}\eta},
\end{align}
where 
 $E_{\bm{k}}^{ \pm}=-t(\cos k_x+\cos k_z) \pm t\sqrt{\cos^2 k_x+\cos^2 k_z}$
are the band energies and $d_{\bm{k}\eta}$ is an operator derived from $c_{\bm{k}\gamma}$ by the unitary transformation. 
\begin{figure}
\onefigure[scale=0.115]{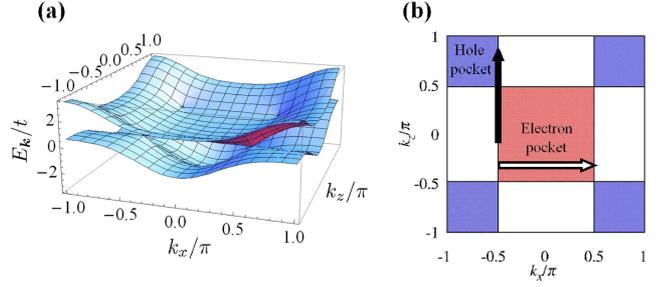}  \caption{(a) The band  structure $E_{\bm{k}}^\pm$  in the first  Brillouin  zone.    The two bands contact with each other at  four points  of $\bm{k}=(\pm\pi/2,\pm\pi/2)$ and $(\pm\pi/2,\mp\pi/2)$.  (b)  Fermi surfaces in the $k_x$-$k_z$  plane.  The red and blue regions are the electron- and hole-pockets, respectively. Black and white arrows represent the inter-band nesting at $\bm{q}=(0,\pi)$ and the intra-band nesting at $\bm{q}=(\pi,0)$ respectively.}
\label{compass_band}
\end{figure}
These band dispersions are shown in fig.~\ref{compass_band}(a). The two bands touch with each other at the four points of  $\bm{k}=(\pm\pi/2,\pm\pi/2)$ and $(\pm\pi/2,\mp\pi/2)$. There is the particle-hole symmetry in the half-filled case. As shown in fig.~\ref{compass_band}(b), the system is a semi-metal, and both the electron- and hole-Fermi surfaces are squares. 
The square-Fermi surfaces are maintained in the bands with an arbitrary value of $\theta$ in eq.~(\ref{sigmaeta}).
Perfect nestings occur at $\bm{q}=(\pi,0),(0,\pi)$  and  $(\pi,\pi)$.

\section{Method}
\begin{figure}
\onefigure[scale=0.14]{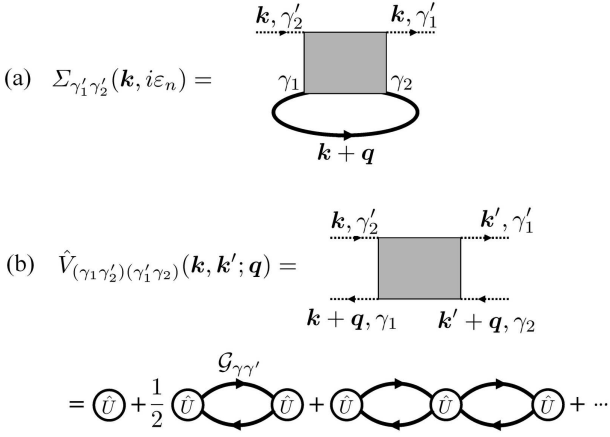}  
\caption{
Diagrammatic representations  of (a) the self energy $\varSigma(\bm{k},i\varepsilon_n)$ and  (b) the effective interaction $\hat{V}(\bm{k},\bm{k}';\bm{q})$. The interaction $\hat{U}$ is  defined in eq.~(\ref{eq:9}). The dotted lines are for the one-particle Green's functions as the external lines, and the  bold lines are for the full one-particle Green's functions.}
\label{Veff}
\end{figure}

The model Hamiltonian in eq.~(\ref{eq:2}) with eqs.~(\ref{eq:2a}) and (\ref{eq:2b}) is analyzed by using  RPA and SC-RPA. We introduce the one-particle Green's functions defined by
\begin{align}
 \mathcal{G}_{\gamma\gamma'}(\bm{k},i\varepsilon_n)&=-\int_0^\beta d\tau
\means{
 c_{\bm{k}\gamma}(\tau)c_{\bm{k}\gamma'}^\dagger}e^{i\varepsilon_n\tau},
\end{align}
where $\varepsilon_n=(2n+1)\pi T$  is the Matsubara frequency, $\tau$ is the imaginary time and $\beta$ is the inverse temperature. The Green's functions at $U=0$, termed $\mathcal{G}^0(\bm{k},i\varepsilon_n)$, are given as 
\begin{align}
 [\mathcal{G}^0(\bm{k},i\varepsilon_n)^{-1}]_{aa}
&=i\varepsilon_n -A_{\bm{k}}^+ E_{\bm{k}}^+-A_{\bm{k}}^-
 E_{\bm{k}}^-+\mu    ,    \label{eq:4}  \\
 [\mathcal{G}^0(\bm{k},i\varepsilon_n)^{-1}]_{bb}
&=i\varepsilon_n -A_{\bm{k}}^+ E_{\bm{k}}^--A_{\bm{k}}^- E_{\bm{k}}^++\mu   , \label{eq:5}\\
 [\mathcal{G}^0(\bm{k},i\varepsilon_n)^{-1}]_{ab}
&= [\mathcal{G}^0(\bm{k},i\varepsilon_n)^{-1}]_{ba} \\ \nonumber 
&=-B_{\bm{k}}( E_{\bm{k}}^+- E_{\bm{k}}^-),  \label{eq:6}
\end{align}
with coefficients 
\begin{equation}
A_{\bm{k}}^{\pm}=\left (1 \mp \cos k_z/ \sqrt{\cos^2 k_x+\cos^2 k_z} \right )/2 ,
\label{eq:ak}
\end{equation}
and 
\begin{equation}
B_{\bm {k}}=-\cos k_x/
\left (2\sqrt{\cos^2 k_x+\cos^2 k_z} \right) . 
\label{eq:bk}
\end{equation}

We also introduce the orbital susceptibility defined as 
\begin{align}
\chi_{(\gamma_1\gamma'_2)(\gamma'_1\gamma_2)}(\bm{q},i\omega_n)
=\int_0^{\beta}\means{\delta 
 n_{ \bm{q}}^{(\gamma_1  \gamma'_2)}(\tau)\delta
 n_{-\bm{q}}^{(\gamma_1' \gamma_2)} }e^{i\omega_n \tau}d\tau , 
\end{align}
with  
$\delta n_{\bm{q}}^{(\gamma \gamma')}=n_{\bm{q}}^{(\gamma \gamma')}
-\means{n_{\bm{q}}^{(\gamma \gamma')}}$ and 
$n_{\bm{q}}^{(\gamma \gamma')}=\sum_{\bm{k}}c_{\bm{k}\gamma}^\dagger  c_{\bm{k}+\bm{q}\gamma'}$.        
From now on, for simplicity, we use the abbreviation for the suffix as 
$\alpha=(\gamma,\gamma')$  which takes  $\uparrow=(a,a)$, $\downarrow=(b,b)$, $+=(a,b)$ and $-=(b,a)$.    
Furthermore, we define the $(zz)$ component of the susceptibility as 
\begin{align}
\chi_{zz}({\bm q}, i \omega_n) =\frac{1}{4} 
\bigl \{ 
 \chi_{\uparrow \uparrow}     ({\bm q}, i \omega_n)
&+\chi_{\downarrow \downarrow} ({\bm q}, i \omega_n)
\nonumber \\
-\chi_{\uparrow \downarrow} ({\bm q}, i \omega_n)
&-\chi_{\downarrow \uparrow} ({\bm q}, i \omega_n)   \bigr \} .  
\label{eq:chizz}
\end{align}

In RPA, the susceptibility is given by
\begin{align}
 \hat{\chi}^{\rm  RPA}&=\hat{\chi}^0 \left (1-
 \hat{U}\hat{\chi}^0 \right)^{-1} ,
 \label{eq:7}
\end{align}
where 
\begin{align}
 \hat{U} =\bordermatrix{ & \uparrow & + & - & \downarrow \cr \uparrow  & 0 & 0 & 0 & -U\cr + & 0 & 0
& U & 0\cr - & 0 & U & 0 & 0\cr \downarrow & -U & 0 & 0 & 0 } , \label{eq:9}
\end{align}
and $\hat{\chi}^0$ is the $4 \times 4$ matrix of $\chi_{\alpha \beta}^0 (\bm{q},i\omega_n)$ which is the bare susceptibility at $U=0$.

In order to consider the band-modification effects, beyond RPA, the susceptibility is calculated by the SC-RPA method, where the one-particle Green's functions and the susceptibilities are calculated self-consistently. The susceptibility is given by the same form with eq.~(\ref{eq:7}) where $\hat{\chi}^0$ is replaced by  $\hat{\bar \chi}^0$ defined by 
\begin{align}
 &\bar{\chi}^0_{(\gamma_1\gamma'_2)(\gamma'_1\gamma_2)}
 (\bm{q},i\omega_n)\nonumber\\&= T\sum_{m}\int_{\rm B.Z.}
\frac{d\bm{k}}{(2\pi)^2}\mathcal{G}_{\gamma'_2\gamma'_1}(\bm{k}+\bm{q},i\varepsilon_m+i\omega_n)
\mathcal{G}_{\gamma_2\gamma_1}(\bm{k},i\varepsilon_m). 
\label{eq:8}
\end{align}
The one-particle Green's functions are obtained by the Dyson's equation $\mathcal{\hat G}=[(\mathcal{\hat G}^0)^{-1}-\hat {\varSigma}]^{-1}$ where the self-energy is given by 
\begin{align}
\varSigma_{\gamma'_1\gamma'_2}(\bm{k},i\varepsilon_n)
=&T\sum_{m}\int_{\rm B.Z.} \frac{d\bm{q}}{(2\pi)^2}
V_{(\gamma_1\gamma'_2)(\gamma'_1\gamma_2)}(\bm{q},i\omega_m)\nonumber\\
&\times \mathcal{G}_{\gamma_2\gamma_1}(\bm{k}+\bm{q},i\varepsilon_n+i\omega_m),\label{eq:3}
\end{align}
as shown in  fig.~\ref{Veff}(a). Here we define the effective interaction $\hat{V}(\bm{q})$ by 
\begin{align}
\hat{V}=(1- \hat{U}\hat {\bar{\chi}}_0)^{-1} \hat{U}
-\frac{1}{2}\hat{U}\hat {\bar{\chi}}_0\hat{U} . 
\label{eq:vv}
\end{align}
Schematic diagrams are shown in fig.~\ref{Veff}(b). The last term in eq.~(\ref{eq:vv}) is required to avoid the double counting of the diagrams. We note that, because of the on-site interaction, $\hat V$ does not depend on both the momenta ${\bm k}$ and ${\bm k'}$. 

\section{Result}
\begin{figure}
\onefigure[scale=0.15]{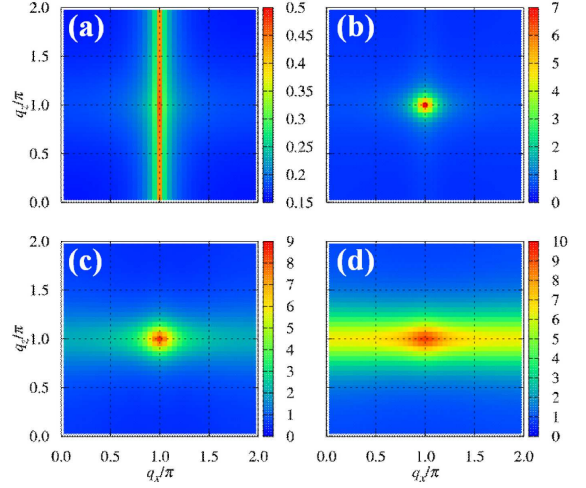}  
\caption{
Contour plots of the $(zz)$ component of the orbital susceptibilities $\chi^{\rm RPA}_{zz}$ as a unit of $t^{-1}$ obtained by the RPA method. Temperature is $T/t=T_c^{\rm RPA}/t+0.01$. Parameter values of $U/t$ are chosen to be (a)0, (b)2, (c)3, and (d)5. 
}
\label{chiRPA}
\end{figure}

First we show the results in RPA. The orbital ordering temperature, $T_c^{\rm RPA}$, is identified as a temperature where the susceptibility diverges  due to the condition of 
\begin{align}
 {\rm Det} \left [1-\hat{U}\hat{\chi}^0(\bm{q},0) \right ]=0.
\label{eq:det0}
\end{align}
The obtained ordering temperature monotonically increases with $U$. The momentum ${\bm q}$, in which the susceptibility diverges at the highest temperature, is $(\pi,\pi)(\equiv {\bm Q})$ for all values of $U$. 

To examine the fluctuation just above $T_c^{\rm RPA}$, the contour maps of the $(zz)$ components of the susceptibilities $\chi^{\rm  RPA}_{zz}({\bm q}, i \omega_n=0)$ for several $U$'s are presented in fig.~\ref{chiRPA}. The temperature is chosen to be $T/t=T_c^{\rm RPA}/t+0.01$ for each value of $U$. 
In the case of small $U$, $\chi_{zz}^{\rm RPA}$ shows large intensity along $(\pi, q_z)$. On the other side, in $U=5$, a large fluctuation emerges along $(q_x, \pi)$. In between, a spot is seen around $(\pi, \pi)$. That is, the anisotropy in $\chi^{\rm RPA}_{zz}$ is changed with the Coulomb interaction. In the RPA scheme, the susceptibility is explicitly given in eq.~(\ref{eq:7}) where $\hat \chi^0$ is independent of $U$. Therefore, the observed characteristic anisotropy in $\chi^{\rm RPA}_{zz}$ is reflected from the $q$ dependence of $\hat \chi^0$ around $T^{\rm RPA}_c$, which depends on $U$. 

In order to understand the origin of the above results, we focus on the bare susceptibility $\chi_{zz}^0({\bm q}, i\omega_n=0)$ at the two representative momenta ${\bm q}=(\pi,0)$ and $(0,\pi)$. In the equation for $\chi_{zz}^0(\bm{q},i\omega_n)$, 
\begin{align}
\chi_{zz}^0(\bm{q},i\omega_n)=\frac{1}{4}\sum_{\eta\eta'}\int_{\rm B.Z.}
\frac{d\bm{k}}{(2\pi)^2}C^{\eta\eta'}_{\bm{k}+\bm{q},\bm{k}}
h_{\bm{k}+\bm{q},\bm{k}}^{\eta\eta'}(i \omega_n) , 
\end{align}
where the coefficient $C^{\eta\eta'}_{\bm{k}+\bm{q},\bm{k}}=A_{\bm{k}+\bm{q}}^\eta A_{\bm{k}}^{\eta'}+A_{\bm{k}+\bm{q}}^{\bar{\eta}} A_{\bm{k}}^{\bar{\eta}'}-2\eta\eta'B_{\bm{k}+\bm{q}}B_{\bm{k}}$ 
and a symbol ${\bar \eta}=\pm$ for $\eta=\mp$. At the momenta ${\bm q}=(\pi, 0)$ and $(0, \pi)$, this coefficient does not depend on ${\bm k}$, and satisfies the relations
\begin{align}
 C^{\eta\eta'}_{\bm{k}+\bm{q},\bm{k}}=\delta_{\eta\eta'}\ \ {\rm at}\ \ \bm{q}=(\pi,0),
 \label{cp0}
\end{align}
and
\begin{align}
 C^{\eta\eta'}_{\bm{k}+\bm{q},\bm{k}}=1-\delta_{\eta\eta'}\ \ {\rm at}\ \ \bm{q}=(0,\pi) . 
\label{c0p}
\end{align}
By using the above relations, we have the equations  
\begin{align}
\chi_{zz}^0(\bm{q}=(\pi,0),0)=\frac{1}{2}\int_{\rm B.Z.} \frac{d\bm{k}}{(2\pi)^2}
h_{\bm{k}+\bm{q},\bm{k}}^{++}(0) , 
\end{align}
and 
\begin{align}
\chi_{zz}^0(\bm{q}=(0, \pi),0)=\frac{1}{2}\int_{\rm B.Z.} \frac{d\bm{k}}{(2\pi)^2}
h_{\bm{k}+\bm{q},\bm{k}}^{+-}(0) , 
\end{align}
where we introduce the Lindhard function 
$h_{\bm{k}+\bm{q},\bm{k}}^{\eta\eta'}(i\omega_n)=[f(E_{\bm{k}+\bm{q}}^\eta)-f(E_{\bm{k}}^{\eta'})]/
(i\omega_n-E_{\bm{k}+\bm{q}}^\eta+E_{\bm{k}}^{\eta'})$, 
the Fermi distribution function 
$f(\varepsilon)=1/[\exp\beta(\varepsilon -\mu)+1]$. 
These relations imply that the intra-band [inter-band] nestings between the vertical Fermi surfaces along $(-\pi/2, k_z)$ and $(\pi/2, k_z)$ are concerned in $\chi_{zz}^0({\bm q}=(\pi, 0), 0)$  [$\chi_{zz}^0({\bm q}=(0,\pi),0)$] (see fig.~\ref{compass_band}), 
and govern the observed anisotropy in the susceptibility.
From now on, 
$\chi_{zz}^0({\bm q}=(\pi, 0), 0)$  and $\chi_{zz}^0({\bm q}=(0,\pi),0)$ 
are termed $\chi_{\rm intra}^0$ and $\chi_{\rm inter}^0$, respectively.

\begin{figure}
\onefigure[scale=0.6]{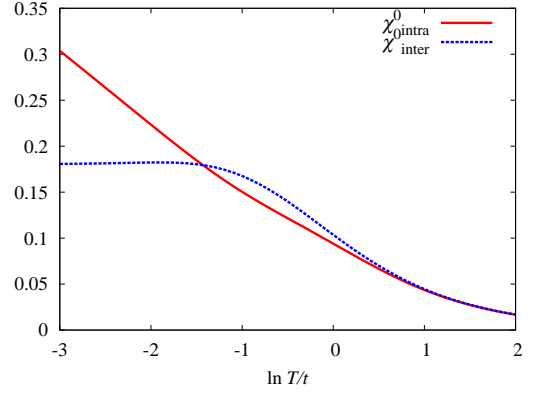} 
\caption{
Temperature dependences of the susceptibilities $\chi_{\rm intra}^0$ and $\chi_{\rm inter}^0$.
}  
\label{chi0}
\end{figure}
\begin{figure}
\onefigure[scale=0.12]{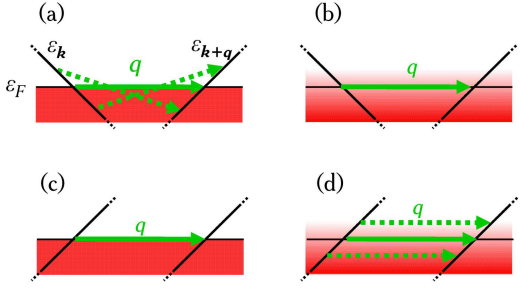}   
\caption{
Schematic views for the band nestings. 
(a) [(b)] is for $\chi^0_{\rm intra}$ in low [high] temperature, 
and (c) [(d)] is for $\chi^0_{\rm inter}$ in low [high] temperature. 
}  
\label{nesting}
\end{figure}
The temperature dependences of  $\chi_{\rm intra}^0$ and $\chi_{\rm inter}^0$ are shown in fig.~\ref{chi0}. 
In low temperatures, $\chi_{\rm  intra}^0$  shows a logarithmic  divergence, and  $\chi_{\rm inter}^0$  is
almost constant.   
There is a crossing point, and $\chi_{\rm inter}^0$ becomes larger than $\chi_{\rm intra}^0$ in high temperatures. 

Schematic intra-band nestings in $\chi_{\rm intra}^0$ are shown in figs.~\ref{nesting}(a) and (b). The two points, $\varepsilon_{\bm{k}}$ and $\varepsilon_{\bm{k}+\bm{q}}$,  connected by ${\bm q}=(\pi, 0)$ 
are located in the left and right branches in the same band in the figure. Band curvatures near the Fermi surface are opposite with each other. In low temperatures [see Fig.~\ref{nesting}(a)], 
there are a number of pairs for  $(\varepsilon_{\bm{k}}, \varepsilon_{\bm{k}+\bm{q}})$ which contributes to $\chi_{\rm intra}^0$ with a fixed momentum ${\bm q}=(\pi, 0)$. With increasing temperature [see Fig.~\ref{nesting}(b)], thermal broadening of the Fermi surfaces reduces the susceptibility, as well known in the nesting in a single-band model.

Inter-band nestings in $\chi_{\rm inter}^0$ 
are also shown in figs.~\ref{nesting}(c) and (d). 
Band curvatures, where $\varepsilon_{\bm{k}}$ and $\varepsilon_{\bm{k}+\bm{q}}$ 
are located, are parallel, in contrast to the case of $\chi_{\rm intra}^0$. 
As a result, in low temperatures,  
a number of pairs for $(\varepsilon_{\bm{k}}, \varepsilon_{\bm{k}+\bm{q}})$,  which satisfy the nesting condition at ${\bm q}=(0, \pi)$, are limited to vicinity of the Fermi surface. 
This is the reason why $\chi_{\rm inter}^0$ is smaller than $\chi_{\rm intra}^0$ in low temperatures. 
With increasing temperature, thermal broadening of the Fermi surfaces  reduces the susceptibilities in both the two cases in $\chi_{\rm inter}^0$  and $\chi_{\rm intra}^0$. 
However, a reduction rate in $\chi_{\rm inter}^0$ is smaller than that in $\chi_{\rm intra}^0$, 
since the nesting conditions in $\chi_{\rm inter}^0$ are satisfied, even when the two points, $\varepsilon_{\bm{k}}$ and $\varepsilon_{\bm{k}+\bm{q}}$, are away from the Fermi surface within an energy range of the order of temperature. 

Now we explain the mechanism of the characteristic $U$ dependence of the anisotropy in $\chi^{\rm RPA}_{zz}$ 
based on the temperature dependences of $\chi_{\rm inter}^0$ and $\chi_{\rm intra}^0$. 
In small $U$, i.e. small $T_c^{\rm RPA}$, the dominant fluctuations are caused by the intra-band nestings.
The susceptibility at $(\pi, 0)$ is larger than that at $(0, \pi)$, because the coefficient 
$C^{\eta\eta'}_{\bm{k}+\bm{q},\bm{k}}$ at ${\bm q}=(\pi,0)$ is finite only for the intra-band fluctuation, i.e. $\eta=\eta'$, as shown in eqs.~(\ref{cp0}) and (\ref{c0p}). 
With increasing the ordering temperature by increasing $U$, 
reduction of the intra-band contribution is more remarkable than 
that of the inter-band one, and the the susceptibility at $(0,\pi)$ becomes larger than that at $(\pi,0)$. 
We conclude that the observed characteristic $U$ dependence of the anisotropy in $\chi^{\rm RPA}_{zz}$ is a consequence of i) the change of the dominant orbital fluctuation from the intra-band nesting to the inter-band one with increasing $T$, and ii) the band- and momentum-dependent coefficient $C^{\eta\eta'}_{\bm{k}+\bm{q},\bm{k}}$. 

\begin{figure}
\onefigure[scale=0.15]{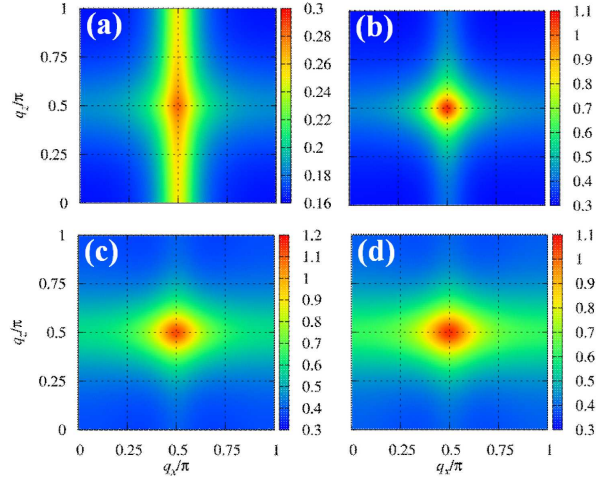}  \caption{
Contour plots of the $(zz)$ components of the orbital susceptibilities 
 as a unit of $t^{-1}$ obtained by the SC-RPA method. 
Temperatures $T/t=0.1$ in all figures. 
Parameter values of $U/t$ are chosen to be (a)0.05, (b)2, (c)3, and (d)3.25.
}  \label{chiSCRPA}
\end{figure}
\begin{figure}
\onefigure[scale=0.11]{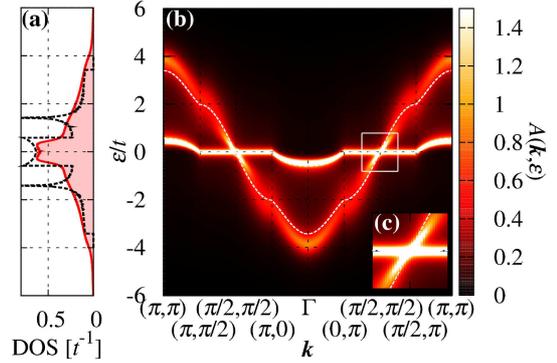} 
\caption{
(a) DOS. (b) a contour map of the one-particle excitation 
spectrum $A(\bm{k},\varepsilon)$ as a unit of $t^{-1}$ in the ${\bm k}$-$\varepsilon$ plane. 
Parameter value of $U/t$ is chosen to be 2 and $T/t=0.1$. 
The dotted line in (a) and that in (b) are for DOS and the band dispersions, respectively, in the case of $U=0$. 
Inset in (b) is an expansion at vicinity of $(\pi/2, \pi/2)$. 
}  
\label{Akw}
\end{figure}
Next we introduce the numerical results obtained by the SC-RPA method.  
The $U$  dependences of the susceptibility  $\chi_{zz}^{\rm   SC-RPA}$ are shown in figs.~\ref{chiSCRPA}. Unlike the results in fig.~\ref{chiRPA}, temperature is fixed at $T=0.1t$.  
In the numerical calculations, a number of meshes in the first Brillouin zone is chosen to be 128$\times$128, 
and that for the imaginary time between $0$ and $\beta$ is chosen to be 1024.
To calculate the excitation spectra, the analytic continuation is adopted as $i\varepsilon_n \rightarrow \varepsilon+i\delta$ with a small constant $\delta=0.01t$. 
At $T=0.1t$, the numerical iterative calculations are converged in the region of $0 \le U \le 3.25$. 
This implies that, at $U=3.25$, the orbital ordering temperature is less than $T=0.1t$ which is lower than that in RPA.  
It is shown in fig.~\ref{chiSCRPA}
that the $U$ dependences of the anisotropy in $\chi_{zz}^{\rm SC-RPA}$ are similar with  
the results in RPA (see fig.~\ref{chiRPA}); with increasing $U$, large fluctuations along $(\pi, q_z)$ is changed into the ones along $(q_x, \pi)$.
 
\begin{figure}
\onefigure[scale=0.55]{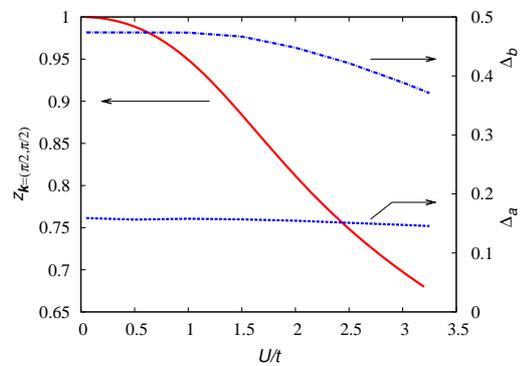} 
\caption{
The renormalization factor at the momentum $\bm{k}=(\pi/2,\pi/2)$ and a number of the states for each orbital near the Fermi surface defined in eq.~(\ref{eq:delta}). 
}
 \label{z0.1}
\end{figure}
The results are interpreted by the one-particle excitation spectrum.
This is defined by $A({\bm k}, \varepsilon)=\sum_\gamma A^{\gamma }({\bm k}, \varepsilon)$ with 
$
A^{\gamma }({\bm k}, \varepsilon)=-(1/\pi){\rm Im} \mathcal{G}_{\gamma \gamma}({\bm k}, i \varepsilon_n \rightarrow \varepsilon+i \delta) .
$
The result of the one-particle spectrum at $U/t=2.0$ is presented in fig.~\ref{Akw}(b), as well as the density of state (DOS) in fig.~\ref{Akw}(a). 
It is seen in the inset that 
the band curvature is strongly reduced from that at $U=0$. 
The mass enhancement is examined from the renormalization factor at the Fermi surface 
defined by 
\begin{align}
z_{\bm{k}}=\frac{1}{2}{\rm    Tr}\left[1-\left.\frac{\partial}     {\partial    \varepsilon}    {\rm
Re}\varSigma(\bm{k},\varepsilon)\right|_{\varepsilon=0}\right]^{-1} , 
\end{align}
where 
values in the two bands are averaged. 
We show in fig.~\ref{z0.1} the $U$ dependence of $z_{\bm{k}}$ at $\bm{k}=(\pi/2,\pi/2)$.
The factor $z_{\rm k}$ monotonically decreases with $U$. 
The results imply that the effective temperature for the low energy fluctuation is enhanced with increasing of $U$ as a result of reduction in the energy scale at vicinity of the Fermi level. 
Therefore, the interpretations for the RPA results introduced previously are also applicable to the present SC-RPA results where the temperature is fixed.

\section{Discussion}

In this section, the numerical results introduced in the previous section are further discussed from the view point of the anisotropic transfer integral  $t_l^{\gamma\gamma'}$ in eqs.~(\ref{eq:tz}) and (\ref{eq:tx}). Let us focus on the $(zz)$ component of the bare orbital susceptibility, $\chi_{zz}^0({\bm q}, 0)$, at the two representative momenta $\bm{q}=(\pi, 0)$ and $(0, \pi)$. As explained in eqs.~(\ref{cp0}) and (\ref{c0p}), it is sufficient to consider the nestings between the vertical Fermi surfaces, i.e. the lines along $(\pi/2, k_z)$ and $(-\pi/2, k_z)$ in fig.~\ref{compass_band}(b). Along these lines, the bare energy bands in eq.~(\ref{eq:bareband}) are given as 
$\varepsilon_{(\pi/2, k_z)}=\varepsilon_{(-\pi/2, k_z)}=-t(1+\sigma^z)\cos k_z$, 
which implies that the $a$ and $b$ orbitals are not hybridized with each other, and the Fermi surfaces, i.e. 
$\varepsilon_{(\pi/2, k_z)}=\varepsilon_{(-\pi/2, k_z)}=0$, 
consist of the flat $b$-orbital band except for the points $(\pm \pi/2, \pm \pi/2)$ and $(\pm \pi/2, \mp \pi/2)$. 
This is a consequence of the anisotropic transfer integral $t_z^{\gamma \gamma'}$; electrons in the $b$ orbital do not hop along the $z$ direction. Therefore, in the sense of the non-interacting electrons, the nestings between the two vertical Fermi surfaces along $(-\pi/2, k_z)$ and $(\pi/2, k_z)$, connected by the momentum ${\bm q}=(\pi, 0)$, are  reflected from the one-dimensional character in the $b$-orbital electrons. 

Now we consider the effect of the Coulomb interaction $U$ on the nestings. This interaction provides the hybridization between the $a$ and $b$ orbitals even on the vertical Fermi surfaces because of the off-diagonal matrix elements e.g. $-U \langle c_b^\dagger c_a \rangle c_a^\dagger c_b$ in the mean-field sense. 
This hybridization induces finite electron motions in the $b$ orbital along the $z$ direction. To check this scenario, we calculate a number of the states for each orbital near the Fermi surface defined by 
\begin{equation}
\Delta_\gamma=\int_{-\varepsilon_c}^{\varepsilon_c}\frac{d\varepsilon}{2\varepsilon_c}
\int_{\rm B.Z.}\frac{d\bm{k}}{(2\pi)^2}A^{\gamma}(\bm{k},\varepsilon), 
\label{eq:delta}
\end{equation}
where $\varepsilon_c$ is the cut-off energy of the order of temperature. The numerical results with $\varepsilon_c/t=0.1$ are presented in fig.~\ref{z0.1}. The interaction reduces more strongly $\Delta_b$ rather than $\Delta_a$; the $b$-orbital character near the Fermi surface is weaken by $U$. This hybridization suppresses the one-dimensional character in the vertical Fermi surfaces, and the intra-band nestings which govern the fluctuation around $(\pi, 0)$ are weaken. In other word, a ratio of the inter-band nesting, which has a dominant  contribution at $(0, \pi)$, to the intra-band one is enhanced by $U$. 

Finally, the observed anisotropy in the orbital fluctuations is discussed with the connection to the original compass model in eq.~(\ref{eq:1}). It is clear from the Hamiltonian in eq.~(\ref{eq:1}) that the correlation between the $z$ component of the pseudo-spins, $T^z$, is stronger along the $z$ direction than that in the $x$ direction. This is caused by the anisotropic exchange interactions originating from the bond-depend transfer integrals in eqs.~(\ref{eq:tz}) and (\ref{eq:tx}) in the perturbational sense. For example, the exchange interaction along the $z$ direction is attributed to the virtual electron hopping in the $a$ orbital along this direction. This fact might correspond to the present observations that $\chi_{zz}({\bm q}, 0)$ has large fluctuation along $(q_x, \pi)$ in the case of large $U$ where the contributions of the $a$ orbital electrons play some crucial role as explained previously. However, in order to examine how the present results in the large $U$ case are connected to the original compass model, further examinations, which are applicable to the intermediate and strong coupling regions, are required. 

In summary, we study the two-dimensional orbital-compass model as an itinerant electronic model. The Hubbard-type Hamiltonian from which the orbital compass model is reproduced is derived. This Hamiltonian is analyzed by the RPA and SC-RPA methods. Rod-like anisotropic fluctuations along $(\pi, q_z)$ in small $U$ is changed into the fluctuations along $(q_x, \pi)$ with increasing $U$. This result originates from the fact that the dominant contributions to the orbital fluctuations are changed from the intra-band nesting to the inter-band one with increasing the interaction. The present study opens a new approach for the orbital compass model.

\acknowledgments
Authors would like to thank M.~Matsumoto and J.~Ohtsuki for the valuable discussions. 
This work was supported by KAKENHI from MEXT, 
Tohoku University ``Evolution'' program, 
and Grand Challenges in Next-Generation Integrated Nanoscience.
JN is supported by the global COE program ``Weaving Science Web beyond Particle-Matter Hierarchy'' of MEXT, Japan.
Parts of the numerical calculations are performed in the supercomputing system in ISSP, the University of Tokyo, and that in YITP, Kyoto University. 

\end{document}